\documentclass[twocolumn,twoside,slac_two]{revtex4}
\usepackage{graphicx}
\usepackage{fancyhdr}
\begin{document}

\title{LHCf Measurements of Very Forward Particles at LHC}

%

\author{T. Sako, for the LHCf Collaboration}
\affiliation{Solar-Terrestrial Environment Laboratory, Nagoya University, Nagoya, Japan}

\begin{abstract}
The LHC forward experiment (LHCf) is specifically designed for measurements 
of the very forward ($\eta$$>$8.4) production cross sections of neutral pions 
and neutrons at Large Hadron Collider (LHC) at CERN.
LHCf started data taking in December 2009, when the LHC started to provide stable 
collisions of protons at $\sqrt{s}$=900\,GeV.
Since March 2010, LHC increased the collision energy up to $\sqrt{s}$=7\,TeV.
By the time of the symposium, LHCf collected 113k events of high energy showers 
(corresponding to $\sim$7M inelastic collisions) at $\sqrt{s}$=900\,GeV and
$\sim$100M showers ($\sim$14 nb$^{-1}$ of integrated luminosity) at $\sqrt{s}$=7\,TeV.
Analysis results with the first limited sample of data demonstrate that LHCf will 
provide crucial data to improve the interaction models to understand very 
high-energy cosmic-ray air showers.
\end{abstract}

\maketitle

\thispagestyle{fancy}


\section{INTRODUCTION}
The development of very  high-energy cosmic-ray observation in the last decade has
dramatically improved the quality of the observation data
\cite{auger-spectrum,auger-anisotropy,auger-Xmax,HiRes}.
However we have not yet obtained consistent interpretation about the nature of the very 
high-energy cosmic-rays.
One of the reasons of this puzzle is the uncertainty of the hadron interaction
at the energy range where we could not test with the man-made accelerators so far. 
Even in the available accelerator energy, we do not have enough data about the very 
forward cross sections that are important to understand the air shower development.
Among hadron collider data, we have only UA7 \cite{UA7} for $\pi^{0}$ at 
$\sqrt{s}$=630\,GeV and ISR data \cite{ISR} for neutrons at $\sqrt{s}$=70\,GeV.

LHC, however, provides us the best opportunity to study hadron interaction at 
$\sqrt{s}$=14\,TeV, corresponding to 10$^{17}$\,eV at the laboratory system. 
LHCf is specifically designed for measurements of the very forward 
($\eta$$>$8.4) production cross sections of neutral pions and neutrons at LHC.
These measurements will set crucial anchor points to constrain the hadron interaction
models to extrapolate into higher energy.
 
In this paper, after introduction of the LHCf experiment and
operation at LHC in 2009 and 2010, some results from the early data are presented.

\section{THE LHCF EXPERIMENT}
The LHCf detectors are installed in the slot of the TANs (Target Neutral Absorber) 
located $\pm$140\,m from the ATLAS interaction point (IP1) and measure the neutral 
particles arriving from the IP.
Inside TAN the beam vacuum chamber makes a Y shaped transition from a single 
common beam tube facing the IP to two separate beam tubes joining to the arcs 
of LHC. 
Charged particles from the IP are swept aside by the inner beam separation 
dipole D1 before reaching the TAN. 
This unique location covers the pseudo-rapidity range from 8.4 to infinity 
(zero degree).

\begin{figure}
  \includegraphics[width=75mm]{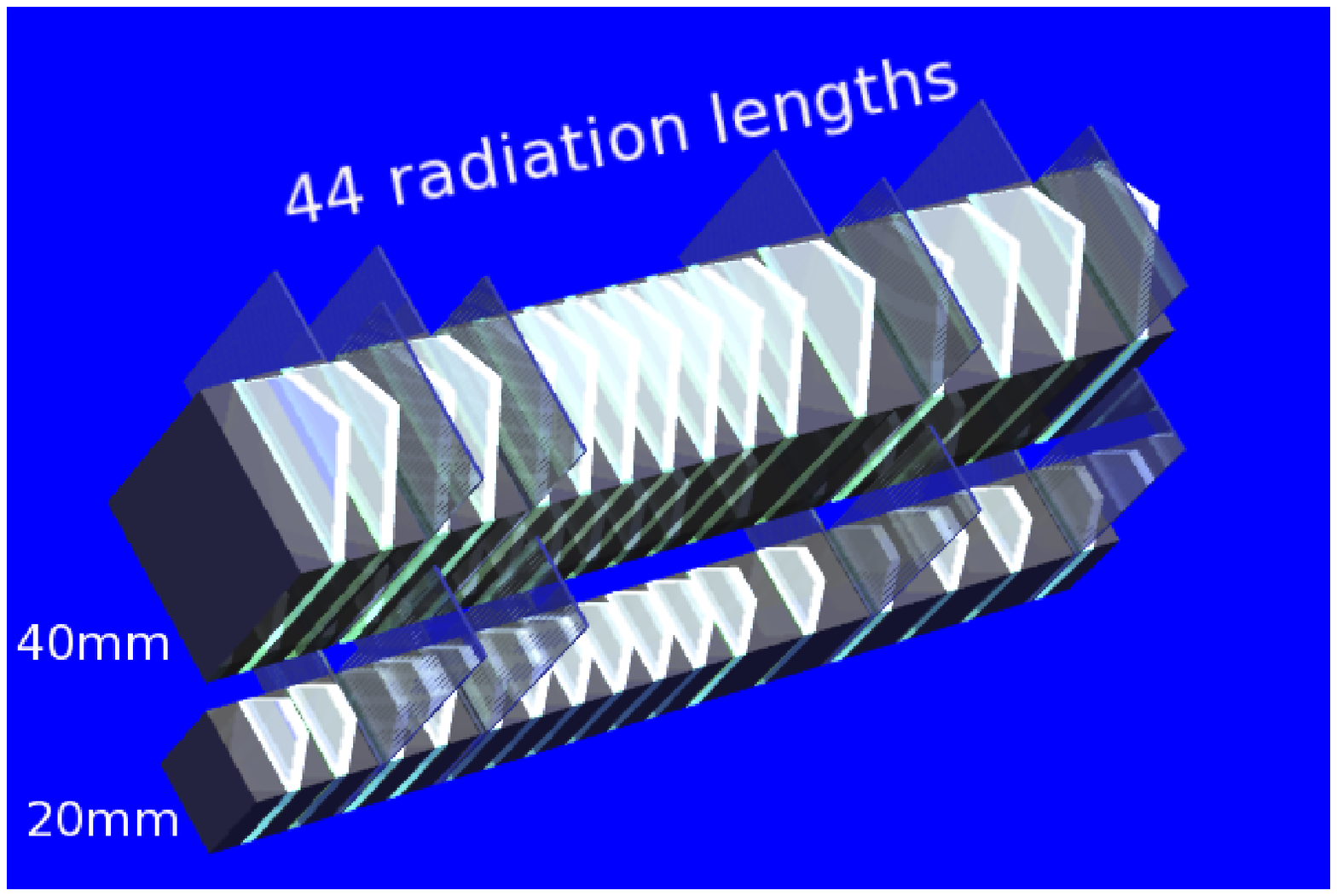}
  \includegraphics[width=75mm]{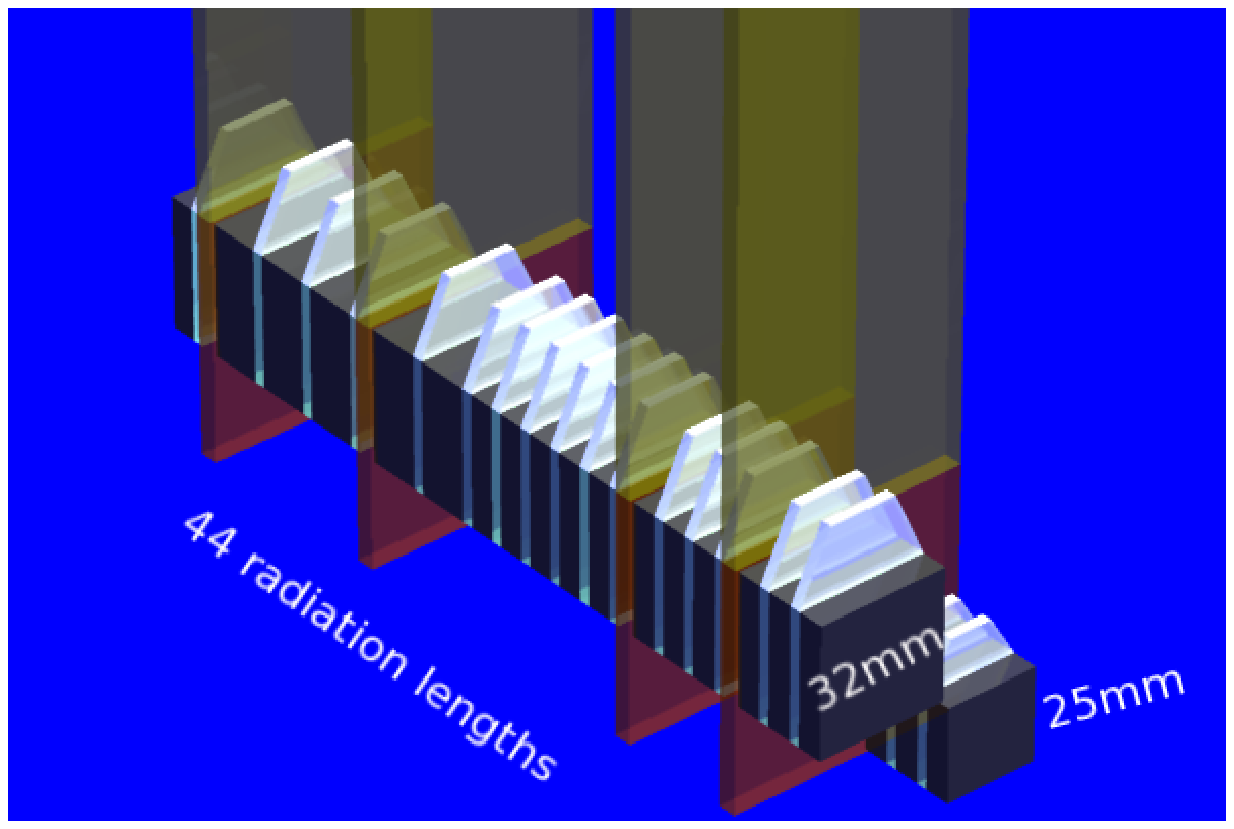}
  \caption{Schematic views of the LHCf detectors (Arm1 and Arm2 in top and 
           bottom, respectively).
           Plastic scintillators (light green) are interleaved with tungsten
           (dark gray) layers.
           Four layers of position sensitive layers (SciFi in Arm1 indicated 
           by light Gray and
           silicon strip detector in Arm2 indicated by brown) are inserted.}
  \label{fig-detector}
\end{figure}

The LHCf detectors are two independent shower calorimeters, named Arm1 and Arm2
installed at the IP8 side and the IP2 side of IP1, respectively. 
Both detectors consist of a pair of small sampling and imaging calorimeters, 
which we call a 'tower' hereafter, made of 16 layers of plastic 
scintillators (3$\,$mm thickness) interleaved with 
tungsten converters (7$\,$mm for the first 11 layers and 14$\,$mm for the rest). 
The longitudinal size is 230$\,$mm or 44 X$_0$ (1.7$\,$$\lambda$) in units of 
radiation length (hadron interaction length).
The transverse sizes of the towers are 20$\,$mm$\times$20$\,$mm and 40$\,$mm$\times$
40$\,$mm in Arm1, and 25$\,$mm$\times$25$\,$mm and 32$\,$mm$\times$ 32$\,$mm 
in Arm2.
Usually the smaller towers cover the zero degree, however, it is adjustable 
using the vertical manipulators.
Four X-Y layers of position sensitive detectors, scintillating fiber (SciFi) 
belts in Arm1 and microstrip silicon sensors in Arm2, are inserted in order to 
provide incident positions of the showers.
The schematic views of the detectors are shown in Figure \ref{fig-detector}.

The calorimeters are designed to have energy and position resolutions better 
than 5$\%$ and 0.2$\,$mm, respectively, for electromagnetic showers with energy
$>$100$\,$GeV. 
Because of the small aperture of the towers, the frequency of multi
particle hits in a single tower is reduced to a reasonable level.
The double tower structure allows us to detect gamma-ray pairs from the decay
of $\pi^{0}$'s with a single gamma-ray induced shower in each tower.
By reconstructing  the invariant mass of gamma-ray pairs, we can identify the
$\pi^{0}$'s and hence measure their energy spectrum.

\begin{figure}
  \includegraphics[width=75mm]{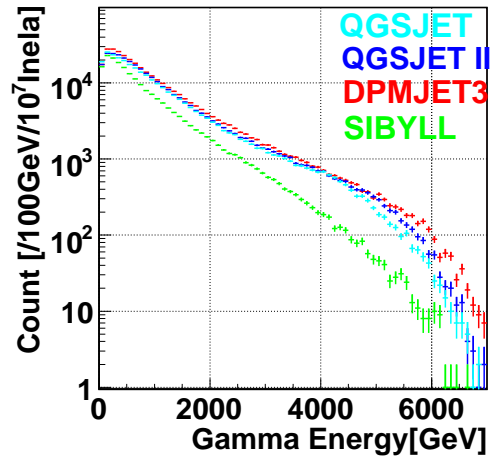}
  \caption{Expected energy spectra of gamma-rays to be measured by LHCf at 
           LHC 14\,TeV collisions.
           Four different models widely used in the cosmic-ray studies are used as
           generators.
           Assumed statistics of 10$^{7}$ inelastic collisions can be achieved with
           a very short operation of LHC even at the commissioning phase.}
  \label{fig-14tevgamma}
\end{figure}

Expected energy spectra of gamma-rays at $\sqrt{s}$=14\,TeV collisions based on 
Monte Carlo (MC) simulation using various interaction models are shown in Figure 
\ref{fig-14tevgamma}.
In this plot, statistical errors are taken into account assuming the total number
of inelastic collisions to be 10$^{7}$ that corresponds to 0.1\,nb$^{-1}$ of
integrated luminosity.
Only short operation of LHCf at the commissioning phase of LHC can provide 
statistically sufficient data to evaluate existing interaction models.

More detail of scientific goal, technical detail and performance of the detectors
are found in
\cite{ref-TDR,ref-JINST,ref-ACTA,ref-Menjo,ref-Silicon,ref-prototype,ref-sps2007}.

\section{OPERATION AT LHC}
LHC has succeeded first physics collisions (stable beams) on 6 December 2009
at $\sqrt{s}$=900\,GeV.
They provided total 0.5M collisions at IP1 in 2009.
After a winter shutdown, LHC succeeded collisions at $\sqrt{s}$=7\,TeV on 30 March
2010 and is gradually increasing the luminosity.
At the time of the ISVHECRI symposium, the integrated luminosity reached at 
14\,nb$^{-1}$. 
Meanwhile they provided 15 times more collisions at $\sqrt{s}$=900\,GeV than 2009.

LHCf has successfully started data taking at the first collision and is accumulating
data at all stable beam condition.
(Note: LHCf has finished operation at this energy at the middle of July 2010 and 
removed the detectors from the LHC tunnel.)
LHCf has accumulated 113k and 100M high energy shower events ($>$10\,GeV
approximately) at 900\,GeV and 7\,TeV collisions, respectively.
The trigger of the LHCf detectors is based on the signals from one of the beam
monitors (BPTX) and existence of high energy shower in any of the calorimeters.
During 2009-2010, LHC has always operated with at least one non crossing bunch 
(having no pair bunch in the other beam) in both beams.
Any high energy particles associated with passage of such bunches at IP1 are
thought to be collision products of beam and residual gas in the beam pipe,
thus background in our measurement.

\section{RESULTS}
\subsection{Data Processing}
In the analyses presented here, we used only early data set.
In the 900\,GeV analyses, we used data taken in 2009, that will increase
about 15 times when we include all data taken in 2010.
In the 7\,TeV analyses, we used data taken only in Mach 2010, that corresponds
to about 1\% of all data taken by the time of the symposium.
(Only for Figure \ref{fig-pizero}, we used more data.)

For each shower, incident position is determined using the information of the
position sensitive layers.
The particles fell within 2\,mm from the calorimeter edge are removed from the
analysis to sustain sufficient energy resolution 
\cite{ref-prototype,ref-sps2007}.
After correcting the non-uniformity of the scintillating photon collection based 
on the position information, the ADC value is converted to the energy deposit and 
summed up over the layers.
Energy of the incident particle is determined from this summation based on the MC
simulation tested below 200\,GeV.
In this energy determination, conversion factor for gamma-rays is applied to all 
particles, therefore the assigned energy so far is 'gamma-ray equivalent energy.'
After the shower leakage correction \cite{ref-prototype}, particles having more 
than 40\,GeV are used in the analyses in the following sections.

For MC simulations, we used various models as generators of the collision particles.
Transport from the collision to the detectors in the beam pipe and the detector
simulation were carried out using the EPICS package \cite{ref-EPICS}.
In the analyses of MC data, same process as data is applied.
In the plots compared with data, the entry of MC simulation is normalized to the
entry of experimental data.

\subsection{Results for 900\,GeV Collisions}
\begin{figure}
  \includegraphics[width=75mm]{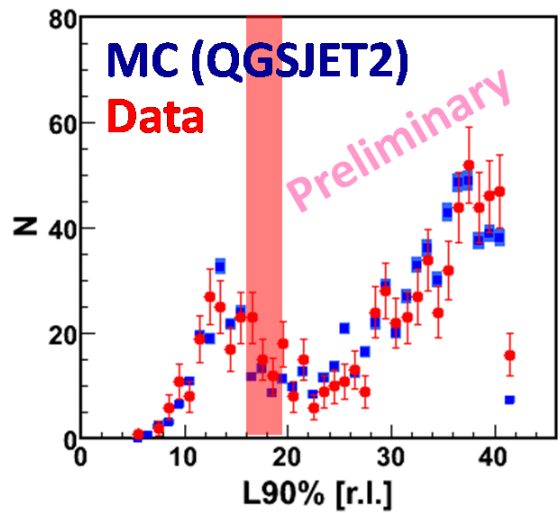}
  \caption{Distribution of the particle ID parameter L$_{90\%}$.
           Red plots are the results of the 900\,GeV collisions while the
           blue rectangles are prediction by MC simulation using QGSJET2
           as a generator.
           A red band from 16 to 20 radiation lengths indicates the criteria
           for particle ID.
           Weak energy dependence (smaller in lower energy, and vice versa)
           is expressed as a band.
           }
  \label{fig-L90}
\end{figure}

According to the models, roughly 50\% of the particles detected by the LHCf 
calorimeters is gamma-rays and the rest is hadrons (mainly neutrons and some 
K$^{0}$ in lower energy).
To identify electromagnetic and hadronic showers, we defined a simple parameter
called 'L$_{90\%}$.'
L$_{90\%}$ is the longitudinal position in radiation length measured from the
entrance of the tower where 90\% of total energy deposit is contained.
Figure \ref{fig-L90} shows the L$_{90\%}$ distributions of the 900\,GeV data and
the MC simulation using QGSJET2 as a generator. 
Clear double peak structure is due to the electromagnetic and hadronic showers.
The peak with smaller (larger) L$_{90\%}$ is caused by electromagnetic (hadronic) 
showers.

From the first look of Figure \ref{fig-L90}, we can expect followings.
Detail study of the L$_{90\%}$ performance and other particle identification methods
are in progress. 
1) MC simulation reasonably reproduces the L$_{90\%}$ distribution.
2) The gamma-ray/hadron ratio predicted by QGSJET2 agrees well with the experimental
result.
3) By cutting at around L$_{90\%}$= 20 r.l., we can obtain best particle 
identification.
In the further analyses, we applied criteria indicated by the red band in the figure.
This is slightly smaller than 20 r.l. to enhance the purity of gamma-ray sample
(actually 90\% purity accoding to the MC simulation) and weak energy dependence is
also introduced.

\begin{figure*}[t]
  \centering
  \includegraphics[width=135mm]{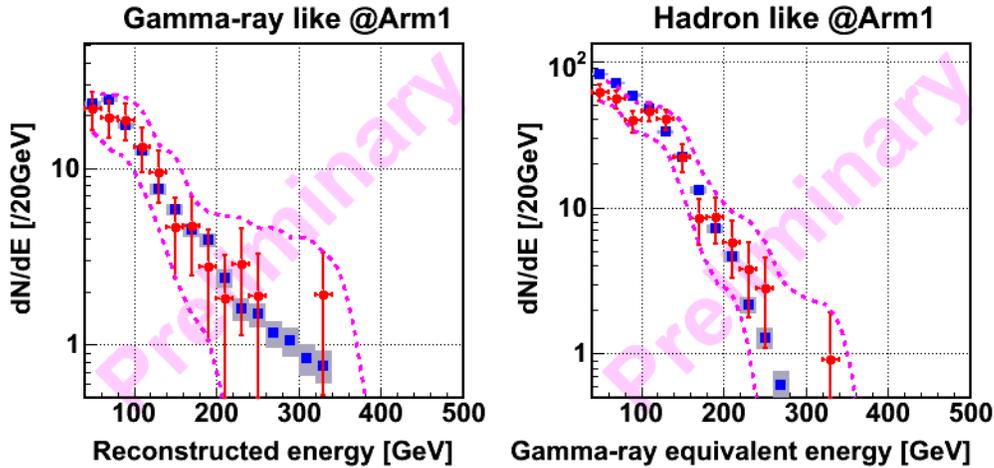}
  \caption{Energy spectra at 900\,GeV collisions measured by the LHCf Arm1 detector
           and prediction by the QGSJET2 model.
           The spectra of gamma-ray (hadron) like events are shown in left (right).
           In the total experimental error shown by the dashed lines, conservative 
           systematic error is included.}
  \label{fig-spectra-900gev}
\end{figure*}

Energy spectra of gamma-ray like and hadron like events after applying the particle
ID criteria are shown in Figure \ref{fig-spectra-900gev}.
The data is from the Arm1 detector after combining the results of two towers.
With this limited statistics, there is no difference between Arm1 and Arm2.
Because of the weak beaming in forward at this lower energy collisions, there are also
no difference in the spectra between small and large towers.

Considering the statistical errors and conservative systematic error at this early
stage, the measured spectra and the prediction by QGSJET2 have a good agreement.

\subsection{Results for 7\,TeV Collisions}
\begin{figure}
  \includegraphics[width=75mm]{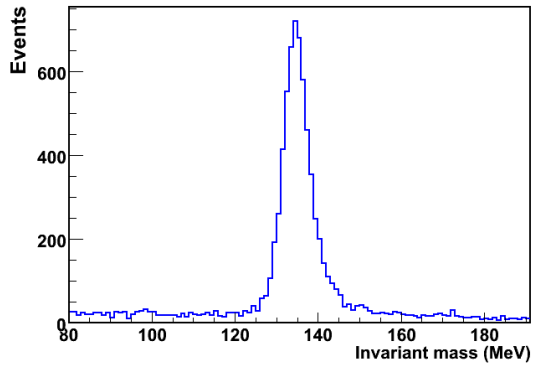}
  \caption{Invariant mass distribution of gamma-ray pairs measured in the Arm2
           detector.
           A clear peak at 135\,MeV is due to the decay of $\pi^{0}$ produced at
           collision and immediately decayed.
           Very low level of combinatorial background is also remarkable.}
  \label{fig-pizero}
\end{figure}

In the analyses of 7\,TeV data, we first applied same particle ID as the 900\,GeV
analysis.
In case of 7\,TeV collisions, gamma-ray pairs decayed from $\pi^{0}$'s produced at 
collisions can hit two towers at the same time.
Using the energy and position information of these gamma-rays and assuming their
parents decayed at IP, the invariant mass of the gamma-ray pairs can be
reconstructed and clear peak at the $\pi^{0}$ mass is found as shown in
Figure \ref{fig-pizero}.
We can see the combinatorial background that makes background in the $\pi^{0}$
analysis is reasonably small.
By selecting the events around the peak, the energy spectrum of $\pi^{0}$'s is
also reconstructed.
We have observed $\pi^{0}$ candidates with energy over 3\,TeV.

\begin{figure*}[t]
  \includegraphics[width=135mm]{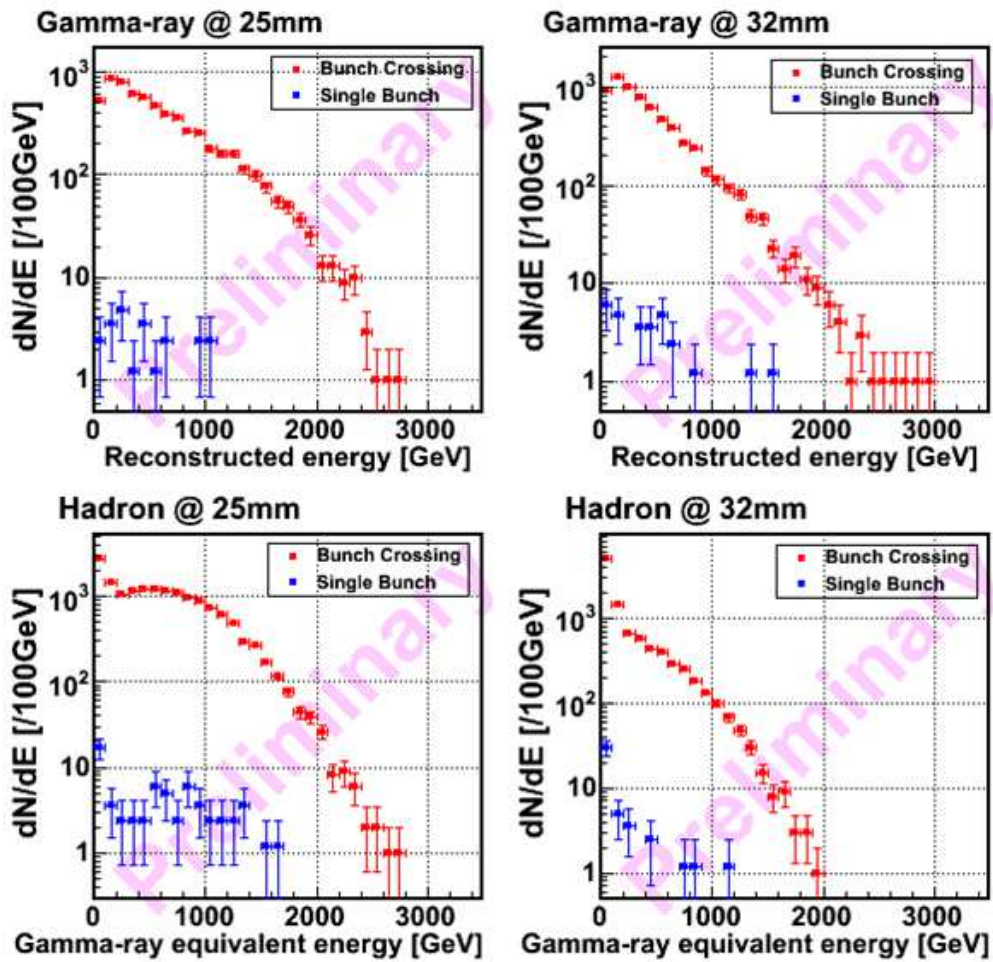}
  \caption{Energy spectra for gamma-ray like and hadron like events measured in the
           Arm2 detector.
           Left plots are the results of the small tower covering zero degree while 
           the right plots are results at non-zero angle measured with the large
           tower.
           Red plots are events associated with the crossing bunched while the
           blue plots are events associated with the non-crossing bunches.}
  \label{fig-spectra-7tev}
\end{figure*}

Energy spectra of gamma-ray like and hadron like events are also shown in
Figure \ref{fig-spectra-7tev}.
Here the spectra measured in the Arm2 detector are separated in the results
of two different towers.
Red and blue markers are events associated with the crossing and non-crossing
bunches, respectively.
From these plots, we can conclude the contamination from the beam-gas background
was 2 orders of magnitude below the signal level and negligible.

When we compare the spectra between small (25\,mm) and large (32\,mm) towers,
harder spectra in the small tower (covering zero degree) are seen both in 
gamma-ray like and hadron like spectra.
This suggests a strong beaming of the high energy particles in very forward
that was not seen in the 900\,GeV data.

\section{FUTURE PLAN}
LHCf plans to finish operation at 7\,TeV collisions before the luminosity reaches
at 10$^{31}$\,cm$^{-2}$s$^{-1}$ because radiation damage to the plastic scintillators
becomes severe and multiple collisions at a single bunch crossing (pile-up) reduce
the quality of data.

\begin{figure}
  \includegraphics[width=75mm]{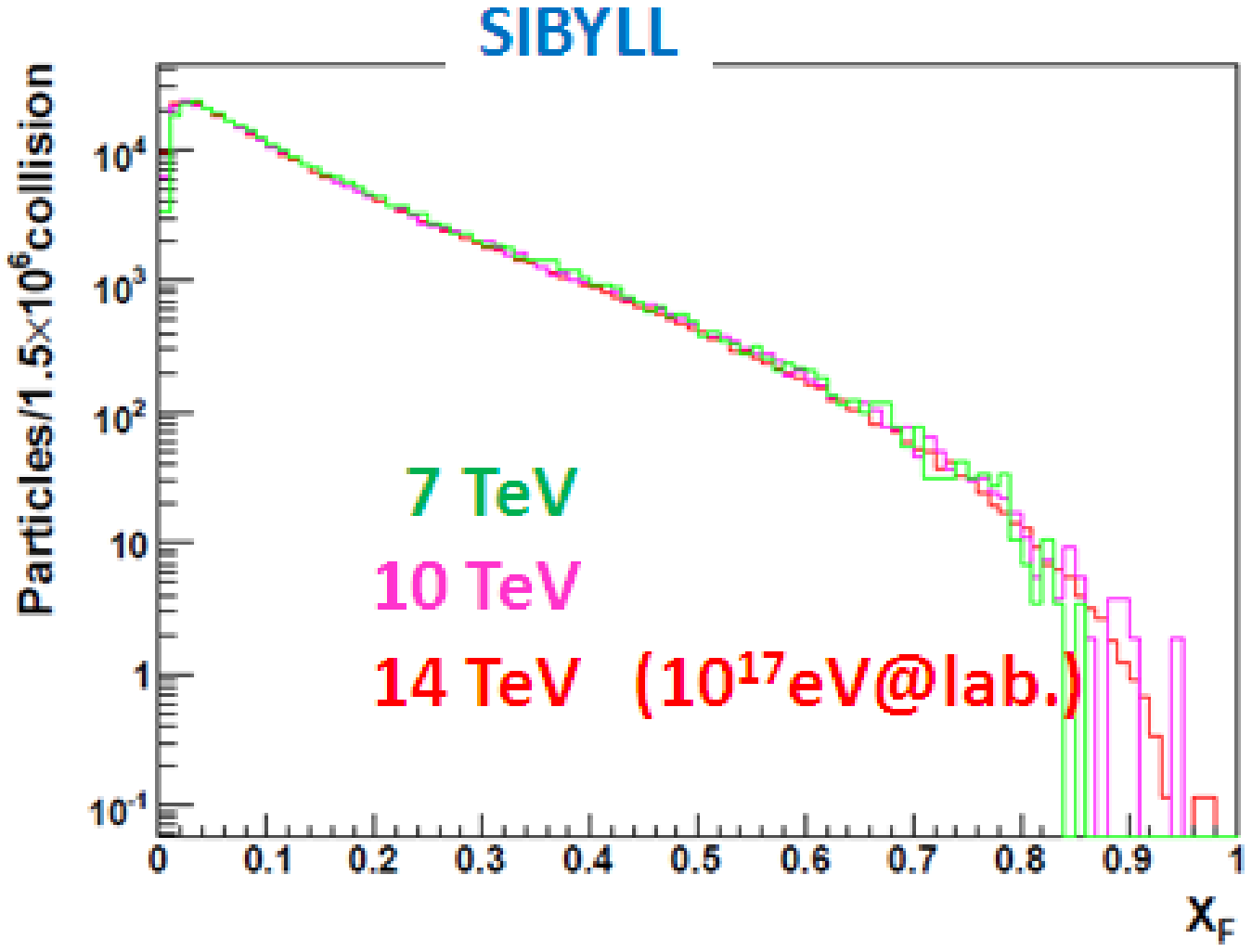}
  \includegraphics[width=75mm]{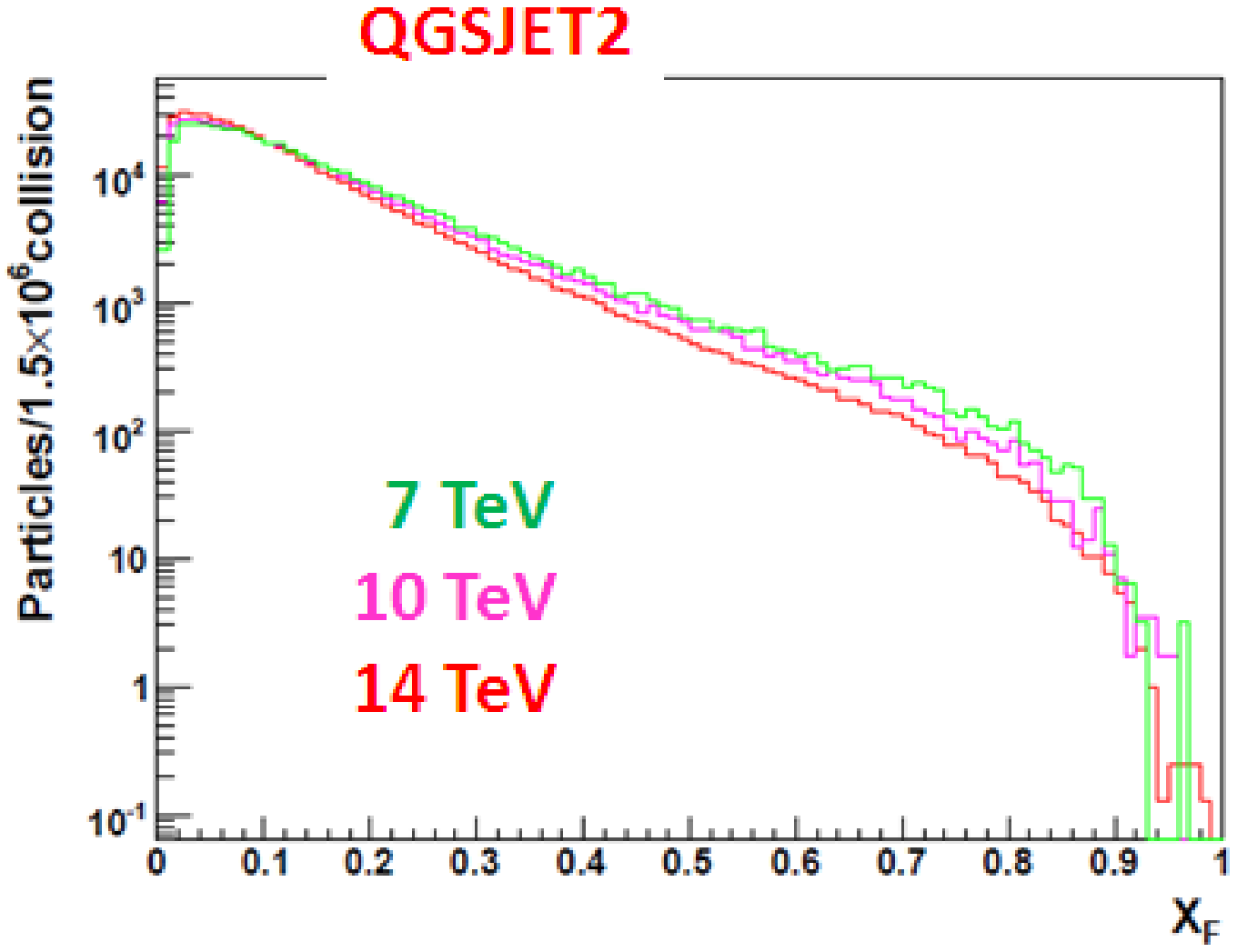}
  \caption{Feynman X distributions (energy spectra normalized by the beam
           energy) of gamma-rays at three different collision energies
           (7\,TeV, 10\,TeV and 14\,TeV) to be measured by the LHCf calorimeters.
           The top and bottom plots are the predictions using the Sybill and
           QGSJET2 generators, respectively.
           Sybill predicts an energy scaling while QGSJET2 predicts softening
           at higher energy.
           Such energy dependence can be tested when LHC increases the beam
           energy in 2013.
           }
  \label{fig-energy_dependence}
\end{figure}

LHC is planning to increase its beam energy up to 7\,TeV (collision energy
of 14\,TeV) in 2013 and LHCf is assured to take data at this energy.
By taking data at three different energies (900\,GeV, 7\,TeV and 14\,TeV), we
can discuss energy dependence of the hadron interaction and it is useful to
extrapolate into the cosmic-ray energy range.
An example of such study is demonstrated in Figure \ref{fig-energy_dependence},
where some expected X$_{F}$ spectra of gamma-rays are plotted for three different
collision energies (7\,TeV, 10\,TeV and 14\,TeV) and two major models (Sybill 
and QGSJET2).
Though Sybill predicts a beautiful scaling over three energies, QGSJET2 predicts
softening of the spectrum.
LHCf will clarify such energy dependence at the operation in 2013.

To be able to fit any commissioning program of LHC at 14\,TeV collisions,
LHCf has started an upgrade work to replace the plastic scintillators with 
crystal scintillators (GSO) those are known to be radiation hard. 

\section{SUMMARY}
LHCf is a dedicated experiment at LHC to measure the cross sections of the
very forward neutral particle emission.
The measurements of LHCf are expected to set a crucial anchor point to 
constrain the hadron interaction models used in the study of cosmic-ray air
showers.
LHCf has successfully started operation when LHC started to provide stable 
collisions in December 2009.
By the time of the symposium, LHC has delivered about 7M inelastic collisions 
at 900\,GeV and 14\,nb$^{-1}$ at 7\,TeV collision at IP1.
LHCf has accumulated data during all these collisions.

In this paper, we presented analysis results using the limited sample of data.
In 900\,GeV analysis, we demonstrated a simple parameter for particle ID, L$_{90\%}$,
reasonably agrees with the prediction of QGSJET2 both in the distribution shape and
gamma-ray/hadron ratio.
Using this particle ID parameter, we derived energy spectra (gamma-ray equivalent)
of gamma-ray like events and hadron like events.
Within large error bars with the limited statistics and conservative systematic
errors, the energy spectra also agree between data and MC.
We will complete the analysis of all data and compare with the other major models.

One of the impressive results in 7\,TeV collisions is a detection of $\pi^{0}$'s.
Thanks to the double tower structure and the excellent energy and position
resolutions, we can reconstruct the invariant mass of gamma-ray pairs. 
In the distribution of the invariant mass, a clear peak at the $\pi^{0}$ mass is
detected.
We have already detected $\pi^{0}$ candidates with energy above 3\,TeV.
The amount and energy spectra of $\pi^{0}$ are essential to understand how much
fraction of hadron interaction is carried into electromagnetic showers.

Energy spectra of 7\,TeV collision data are also presented only using 1\% of
total data set.
We could demonstrate, using the crossing and non-crossing bunch configurations
at LHC, that the contamination from the collisions between beam and residual 
gas is two orders of magnitude below the beam-beam collisions. 
The comparison of spectra between towers shows strong beaming (hard 
spectrum) in the small tower covering zero degree.

LHCf will finish operation when the luminosity reaches at 
10$^{31}$\,cm$^{-2}$s$^{-1}$ to avoid severe radiation damage and pile-up
events.
[Detectors were removed on 20-July 2010.]
LHCf will come back when LHC increases the beam energy up to 7\,TeV (collision
energy of 14\,TeV) planned in 2013.
The data of LHCf at three (or more at intermediate energies) different energies
will be very useful to construct future hadron interaction models those are
used in the study of cosmic-ray air showers at much higher energy than LHC.

\bigskip 
\begin{acknowledgments}
We thank to all the staff in CERN, especially staff to operate the accelerator
complex for providing excellent beams.
We also thank to the ATLAS collaboration to provide us various infrastructure
at IP1.
The experiment is supported by Grant-in-Aid (Kakenhi) by the Ministry of 
Education, Culture, Sports, and Technology (MEXT) of Japan and by Istituto
Nazionale di Fisica Nucleare (INFN) in Italy.
\end{acknowledgments}

\bigskip 

\begin{thebibliography}{99} 


\bibitem{auger-spectrum} J. Abraham et al., Phys. Rev. Lett., 101 (2008) 061101
\bibitem{auger-anisotropy}  The Pierre Auger Collaboration, Science, 318 (2007) 938-943
\bibitem{auger-Xmax} J. Abraham et al., Phys. Rev. Lett., 104 (2010) 091101
\bibitem{HiRes} R.U. Abbasi, et al., Phys. Rev. Lett., 104 (2010) 161101
\bibitem{UA7} E. Par\'{e} et al., Phys. Lett. B, 242 (1990) 531-535
\bibitem{ISR} W. Flauger and F. M\"{o}nnig, Nucl. Phys. B, 109 (1976) 347-356
\bibitem{ref-TDR} LHCf Technical Design Report, CERN-LHCC-2006-004
\bibitem{ref-JINST} O. Adriani et al., JINST, 3 (2008) S08006
\bibitem{ref-ACTA} R. D'Alessandro et al., Acta. Physica, Polonica B, 38 (2007) 829
\bibitem{ref-Menjo} H. Menjo et al., Astropart. Phys. (2009) submitted
\bibitem{ref-Silicon} O. Adriani et al., JINST, 5 (2010) P01012
\bibitem{ref-prototype} T. Sako, et al., NIM, A578 (2007) 146
\bibitem{ref-sps2007} O. Adriani, et al., in preparation
\bibitem{ref-EPICS} K. Kasahara, EPICS web page, http://cosmos.n.kanagawa-u.ac.jp/
\

\end{thebibliography}

\end{document}